\documentclass[sigconf,screen,nonacm]{acmart}
\AtBeginDocument{%
  }

\setcopyright{none}
\usepackage{amsmath}
\usepackage{amsthm}
\usepackage{mathtools}
\usepackage{booktabs}
\usepackage{array}
\usepackage{multirow}
\usepackage{xcolor}
\usepackage{placeins}
\usepackage{algorithm}
\usepackage{algpseudocode}
\AtBeginDocument{%
  }

\begin{document}

\title{SourceFlow: Collaborative Source Coupling for Flow-Based Recommendation}

\author{Chenghao Wu}
\affiliation{%
  \institution{The Hong Kong University of Science and Technology (Guangzhou)}
  \city{Guangzhou}
  \country{China}}
\affiliation{%
  \institution{Tencent Inc.}
  \city{Shenzhen}
  \country{China}}
\email{cwu319@connect.hkust-gz.edu.cn}

\author{Yongqi Zhou}
\affiliation{%
  \institution{Tencent Inc.}
  \city{Shenzhen}
  \country{China}}

\author{Jin Chen}
\affiliation{%
  \institution{Tencent Inc.}
  \city{Shenzhen}
  \country{China}}

\author{Yifeng Zhou}
\affiliation{%
  \institution{Tencent Inc.}
  \city{Shenzhen}
  \country{China}}

\author{Jinchang Zhu}
\affiliation{%
  \institution{The Hong Kong University of Science and Technology (Guangzhou)}
  \city{Guangzhou}
  \country{China}}

\author{Junwei Pan}
\affiliation{%
  \institution{Tencent Inc.}
  \city{Shenzhen}
  \country{China}}

\author{Chao Zhou}
\affiliation{%
  \institution{Tencent Inc.}
  \city{Shenzhen}
  \country{China}}

\author{Zhixiang Feng}
\affiliation{%
  \institution{Tencent Inc.}
  \city{Shenzhen}
  \country{China}}

\author{Chengguo Yin}
\affiliation{%
  \institution{Tencent Inc.}
  \city{Shenzhen}
  \country{China}}

\author{Menglin Yang}
\affiliation{%
  \institution{The Hong Kong University of Science and Technology (Guangzhou)}
  \city{Guangzhou}
  \country{China}}
\email{menglinyang@hkust-gz.edu.cn}

\begin{abstract}
  Diffusion-based recommenders learn user preferences by denoising interaction
  histories, but the added noise can distort the original preference signals.
  Flow-based collaborative filtering (CF) instead learns a discrete
  path from a sampled source to the user's observed training
  interactions (the target history), enabling inference in a few steps.
  Existing flow-based CF methods sample sources using global item
  frequencies to reflect the sparsity and popularity of observed interactions.
  Each source is sampled independently of its target history, leaving the
  predictor without user-specific source cues for learning to reconstruct that
  history.
  We propose SourceFlow, which brings collaborative structure into the source
  distribution itself rather than adding further conditioning to the predictor.
  Like inference from observed histories, training uses collaborative evidence
  to predict missing interactions.
  Specifically, Neighborhood Coupling constructs displaced binary sources by
  replacing historical items with unobserved collaborative alternatives.
  Collaborative Consensus further refines replacement probabilities using
  evidence from other interactions of users who share the observed item.
  Stretched Restoration retains these replacements at late flow times during training,
  providing more opportunities to recover the original interactions.
  Experiments on five recommendation datasets show consistent improvements over competitive
  baselines, including a mean relative Recall@10 gain of 15.7\% over the
  strongest flow-based baseline. Controlled experiments confirm that these gains
  depend on correctly pairing collaborative sources with their target histories,
  as reassigning the same source vectors across users reverses the gains.
  Code will be publicly released upon acceptance.
\end{abstract}



\makeatletter
\gdef\shortauthors{Wu et al.}
\def\@mkauthors{%
  \global\setbox\mktitle@bx=\vbox{\noindent
    \unvbox\mktitle@bx\par\medskip
    \centering\large
    Chenghao Wu\textsuperscript{1,2},
    Yongqi Zhou\textsuperscript{2},
    Jin Chen\textsuperscript{2},
    Yifeng Zhou\textsuperscript{2},
    Jinchang Zhu\textsuperscript{1},
    Junwei Pan\textsuperscript{2},
    Chao Zhou\textsuperscript{2},
    Zhixiang Feng\textsuperscript{2},
    Chengguo Yin\textsuperscript{2},
    Menglin Yang\textsuperscript{1}\textsuperscript{\textdagger}\par
    \smallskip
    {\small
      cwu319@connect.hkust-gz.edu.cn,
      menglinyang@hkust-gz.edu.cn\textsuperscript{\textdagger}\par}
    \smallskip
    {\small
      \textsuperscript{1}The Hong Kong University of Science and Technology (Guangzhou)\qquad
      \textsuperscript{2}Tencent Inc.\par}
    \bigskip}}
\makeatother
\maketitle
\begingroup
\renewcommand{\thefootnote}{\textdagger}%
\footnotetext[1]{Corresponding author.}
\endgroup

\begin{figure}[t]
  \centering
  \scalebox{1}[1.1025]{\includegraphics[trim=5.7bp 0bp 5.7bp 0bp,clip,width=\columnwidth]{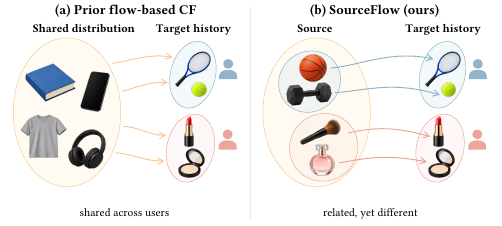}}
  \caption{
Illustration of source--target coupling in flow-based CF training. (a) Prior methods (FlowCF, CoFlowCF) sample sources from a shared global distribution. (b) SourceFlow (Ours) replaces observed items with unobserved collaborative alternatives, retaining collaborative preference structure.
  }
  \Description{Product images illustrate source items and target histories.
  Blue and coral indicate users with sports and beauty preferences. The left
  panel connects a shared global distribution to target histories. The right
  panel connects unobserved collaborative alternatives to their paired target
  items. Arrows illustrate the training reconstruction task.
  The positions and shaded regions are conceptual.}
  \label{fig:collaborative-source-intuition}
  \vspace{-1\baselineskip}
\end{figure}

\section{Introduction}
\begin{figure*}[t]
  \centering
  \begin{minipage}[t]{0.57\textwidth}
    \vspace{0pt}
    \centering
    {\small\textbf{(a)} Training-state construction\par}
    \vspace{1mm}
    \includegraphics[width=\linewidth]{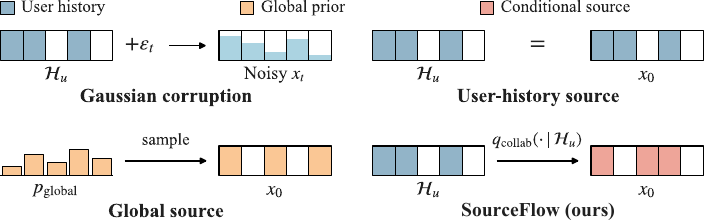}
  \end{minipage}\hfill
  \begin{minipage}[t]{0.41\textwidth}
    \vspace{0pt}
    \centering
    {\small\textbf{(b)} Property comparison\par}
    \vspace{1mm}
    \begingroup
    \setlength{\tabcolsep}{3.2pt}
    \renewcommand{\arraystretch}{1.62}
    \newcommand{\figyes}{\textcolor{green!55!black}{$\checkmark$}}
    \newcommand{\figno}{\textcolor{red!70!black}{$\times$}}
    \resizebox{\linewidth}{!}{%
      \begin{tabular}{@{}lccc@{}}
        \toprule
        \textbf{Method} &
        \shortstack{\textbf{\strut Binary-state}\\\textbf{\strut preserving}} &
        \shortstack{\textbf{\strut Non-trivial}\\\textbf{\strut task}} &
        \shortstack{\textbf{\strut User-preference}\\\textbf{\strut cues}} \\
        \midrule
        Gaussian corruption  & \figno  & \figyes & \figyes \\
        User-history source  & \figyes & \figno  & \figyes \\
        Global source        & \figyes & \figyes & \figno  \\
        \midrule
        \textbf{SourceFlow (ours)} & \figyes & \figyes & \figyes \\
        \bottomrule
      \end{tabular}%
    }
    \endgroup
  \end{minipage}
  \caption{Comparison of source choices for flow-based recommendation.
  Copying the history leaves no reconstruction task. Gaussian corruption produces
  continuous states, while global sources are sampled independently of the
  paired history. SourceFlow combines binary states, a non-trivial reconstruction
  task, and user-preference cues by constructing displaced binary sources
  conditioned on the paired user's history.}
  \Description{A two-panel comparison of training-state construction and source
  properties. The left panel schematically shows Gaussian corruption, a
  user-history source, a global source, and the conditional SourceFlow source.
  The right panel compares the same four methods by binary-state preservation,
  a non-trivial task, and user-preference cues. SourceFlow satisfies all three properties.}
  \label{fig:training-state-source-comparison}
\end{figure*}

Personalized recommendation plays an important role in content discovery on Web platforms,
where users with diverse interests browse evolving item collections \citep{covington2016deep,gomez2015netflix}. On e-commerce and
review platforms, recommendation systems use recorded user--item interactions
to rank products and services according to individual
preferences~\cite{li2023diffurec,yang2023generate,hu2024generate}.
Collaborative filtering (CF) supports this task by inferring user preferences from collective
interaction patterns \citep{sedhain2015autorec, salakhutdinov2007restricted}. Learning such
preferences remains challenging when implicit feedback is sparse and noisy.
Generative CF addresses this problem by modeling user interaction
distributions, with methods ranging from variational autoencoders to 
diffusion models \citep{askari2021variational, wang2023diffusion, li2024recdiff}. DiffRec, for example, reconstructs user histories through
iterative denoising~\cite{wang2023diffusion,wei2025diffusion}.
Gaussian corruption converts binary interaction vectors into continuous states.
Such continuous values cannot directly signal the occurrence of an interaction,
as binary implicit feedback does~\cite{liu2025flow,hu2008collaborative,he2017neural,pan2008one}.

Flow matching offers an alternative by learning a time-dependent vector field
that transports an explicit source distribution toward the target
distribution. FlowCF brings this framework to collaborative filtering through a
behavior-guided Bernoulli prior and discrete interpolation, which preserves binary
interaction states while enabling few-step inference~\cite{liu2025flow}.
CoFlowCF further introduces contrastive supervision over
hidden representations to improve user-level discrimination while capturing
shared interests~\cite{jin2026contrastive}.


FlowCF and CoFlowCF sample training sources from global item frequencies to
reflect interaction sparsity and popularity~\cite{liu2025flow,jin2026contrastive}.
At inference, however, these models start from a user's observed history and
rely on its preference cues to rank unobserved items. During training, sources
are sampled independently of the target history (the paired user's observed
training history to be reconstructed). These sources therefore provide no
user-specific cues for reconstruction. This raises a central question: how can
source design help flow-based CF learn user preferences more effectively?

Figure~\ref{fig:training-state-source-comparison} illustrates the requirements
for this source design. Copying the target history leaves no reconstruction task.
Gaussian perturbations create a reconstruction task by adding noise to binary
histories. The continuous states can obscure the preference
signals~\cite{wang2023diffusion}. A suitable source should be binary, make reconstruction non-trivial, and retain user-specific collaborative cues for reconstruction.

We propose \textbf{SourceFlow} to reshape the reconstruction task through source
design rather than compensate for missing user-specific source cues with
additional predictor conditioning. SourceFlow replaces observed items in the
paired user's history with unobserved collaborative alternatives from the
training interaction graph. The resulting sources share no items
with their target histories, yet retain the paired user's collaborative
preference structure for reconstruction
(Figure~\ref{fig:collaborative-source-intuition}). Just as inference starts
from a user's observed history to rank unobserved items, training starts from
user-specific collaborative cues to reconstruct the target history.

SourceFlow comprises three components. \textbf{Neighborhood Coupling}
treats each historical item as an anchor and samples an unobserved replacement
from its collaborative neighborhood.
\textbf{Collaborative Consensus} reweights the same candidates using support
from other interactions of users who interacted with the anchor item.
Broader collaborative support complements direct item co-occurrence.
\textbf{Stretched Restoration} slows the decay of selected replacements along
the training path. This keeps more historical items replaced at late flow times
and gives the predictor more opportunities to recover them from collaborative
cues. SourceFlow changes only the construction and interpolation of training
states. The scoring network and inference procedure remain unchanged.

\begingroup
Experiments on five recommendation datasets demonstrate consistent
improvements over competitive baselines, including a mean relative Recall@10
gain of \textbf{15.7\%} over the strongest flow-based baseline. Ablations show
consistent improvements from collaborative source design, with gains beyond
those achieved by uniform or popularity-based replacement. Collaborative
sources retain preference-group structure after
item replacement, and trained
predictors favor their target histories over other users' histories in
reconstruction. Reassigning the same source vectors across users removes the
recommendation gains. The benefit therefore depends on pairing collaborative
alternatives with the histories that guided their selection. Our contributions
are threefold:
\begin{itemize}
\item We identify and formalize conditional source coupling as a design choice
  for flow-based CF training, showing how it changes the
  conditional reconstruction problem with target histories and the objective
  held fixed.
\item We propose SourceFlow, which constructs displaced binary sources from
  collaborative alternatives and retains these alternatives at late flow times.
  This trains the predictor to infer preferences across items by recovering
  missing interactions from user-specific collaborative cues.
\item We demonstrate consistent improvements on five datasets and show that
  collaboratively coupled sources preserve user-related preference structure
  while improving reconstruction and recommendation.
\end{itemize}

\section{Related Work}
\subsection{Generative Collaborative Filtering}
Generative CF learns user preferences by reconstructing interaction histories
and uses the resulting item scores for recommendation~\cite{lee2025collaborative}. CDAE reconstructs
histories from corrupted inputs~\cite{wu2016collaborative}, while variational
models such as Mult-VAE, MacridVAE, and RecVAE learn latent preference
representations for
reconstruction~\cite{liang2018variational,ma2019learning,shenbin2020recvae}.

Diffusion-based CF trains models to reconstruct histories from states at
different perturbation levels. The perturbation creates a
reconstruction task, but excessive noise can erase personalized information.
DiffRec therefore limits Gaussian corruption to retain preference signals in
the perturbed history~\cite{wang2023diffusion}. Collaborative structure can
also inform the reconstruction process. GiffCF constructs perturbed states by
smoothing interaction histories over an item--item graph, yielding continuous
preference signals~\cite{zhu2024graph}. CF-Diff supplies
multi-hop collaborative connectivity as a condition for denoising, providing
information beyond the user's direct
interactions~\cite{hou2024collaborative}. These diffusion models reconstruct
histories from continuous states, with user information carried by the
perturbed input or supplied separately as a denoising condition.

\subsection{Flow-based Recommendation}
Flow matching offers deterministic transport through a learned velocity field
along an explicit source--target path. FlowCF adapts this formulation
to collaborative filtering with a Bernoulli source parameterized by global
item frequencies and a mask-based interpolation whose states remain binary,
enabling few-step reconstruction of user histories~\cite{liu2025flow}.
CoFlowCF retains the same source and discrete path, but augments the flow
encoder with dropout-based contrastive views to improve inter-user
discrimination without directly separating scores for shared
items~\cite{jin2026contrastive}. Related flow models have also appeared in
sequential recommendation. FMRec uses straight trajectories and deterministic
Euler sampling for target-item generation~\cite{liu2025fmrec}. FAVE replaces
Gaussian noise in its consolidation stage with a semantic anchor derived from
the masked interaction history, then learns an average velocity with a
straightness constraint for one-step prediction~\cite{shi2026fave}.

\subsection{Source--Target Coupling in Flow Matching}
Beyond recommendation, flow matching commonly trains on independently paired samples from a simple
source and the data distribution~\cite{lipman2022flow}. Changing this
joint coupling can simplify the regression target without changing either
endpoint marginal \citep{liu2022fast, albergo2022building,zhang2025hierarchical}. OT-CFM constructs minibatch optimal-transport pairings to
obtain simpler flows, while Multisample Flow Matching generalizes such
pairings and shows lower gradient variance, straighter flows, and stronger
few-step generation~\cite{tong2023improving,pooladian2023multisample, kim2025simple}.
Stochastic interpolants with data-dependent couplings directly sample the base
conditional on a target observation, enabling conditional transport under the
same squared-error objective~\cite{albergo2023stochastic}. For conditional
generation, C$^2$OT adds condition-aware transport costs to avoid the
train--test mismatch induced by condition-agnostic pairings~\cite{cheng2025curse}.
A complementary line changes the source distribution itself. Conditional-prior
flow matching centers a source distribution near a target-space point
predicted from the condition~\cite{issachar2025designing}, whereas
CSFM learns a condition-dependent source and stabilizes it through variance
regularization and directional
alignment~\cite{kim2026better}.

These studies examine source design and coupling in continuous spaces. In
binary CF, constructing a source requires selecting discrete items, but proximity
between binary interaction vectors does not directly encode collaborative relations between different
items. Diffusion
recommenders preserve user information through history-derived corruption or
conditioning. FlowCF and CoFlowCF preserve binary states and efficient
inference, but draw each training source from the same global distribution
without using the paired user's history. SourceFlow instead
constructs a displaced binary source using collaborative alternatives that
retain the user's preference cues.

\par
\endgroup

\section{Preliminaries}
\label{sec:preliminaries}
\subsection{Recommendation Task}
Let $\mathcal U$ and $\mathcal I$ denote the user and item sets. Let
$\mathcal H_u$ contain user $u$'s observed training items. We represent this history as
$\mathbf x_{1,u}=\mathbf b(\mathcal H_u)\in\{0,1\}^{|\mathcal I|}$,
where $\mathbf b(\cdot)$ converts an item set into its binary indicator.
The recommendation task is to rank unobserved items according to the user's
preferences.
Generative CF treats observed histories as samples from a distribution of
user behavior. We denote the distribution of target histories by
$p_{\mathrm{data}}$. By reconstructing histories sampled from this distribution,
the model learns interaction patterns across users. Its predicted item scores
are used to rank unobserved items.

\begin{figure}[t]
  \centering
  \includegraphics[width=\columnwidth]{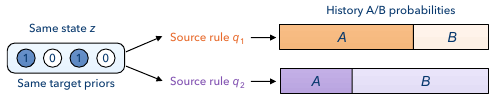}
  \caption{For the same intermediate state and target distribution, changing the
  source rule can alter the posterior over target histories.}
  \Description{A fixed binary intermediate state is evaluated under a fixed
  target distribution and two different source rules. The two rules assign different
  posterior proportions to candidate target histories A and B.}
  \label{fig:source-rule-posterior}
\end{figure}

\begin{figure*}[t]
  \centering
  \includegraphics[width=\textwidth]{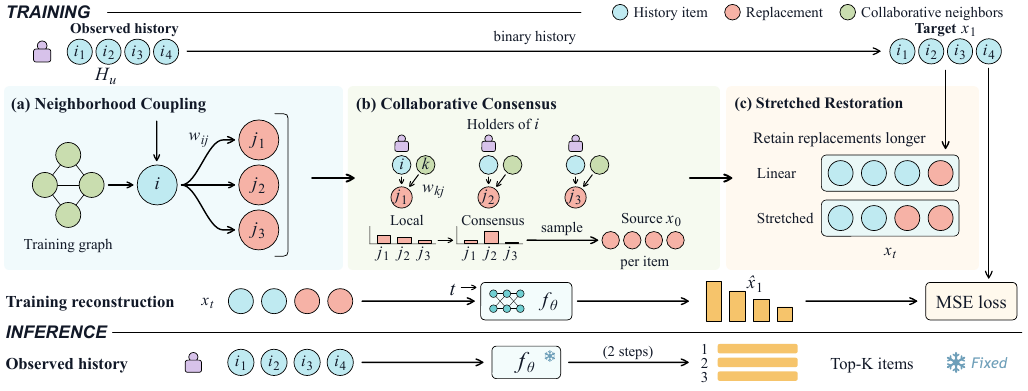}
  \caption{Overview of SourceFlow: collaborative source construction and Stretched
  Restoration during training, with inference from observed user histories.}
  \Description{The training pipeline combines Neighborhood Coupling,
  Collaborative Consensus, and Stretched Restoration to reconstruct user
  histories. The inference pipeline starts from observed histories and uses
  two prediction steps to rank items.}
  \label{fig:sourceflow-overview}
\end{figure*}

\subsection{Flow Matching}
Flow matching learns a time-dependent velocity field
$\mathbf v_\theta(\mathbf x,t)$ that transports a source distribution $p_0$
toward $p_{\mathrm{data}}$. For CF training, source samples provide starting
interaction states, and target samples are users' training histories.
The state at time $t\in[0,1]$ evolves according to the ordinary differential equation
\begin{equation}
  \frac{\mathrm d\mathbf x_t}{\mathrm dt}
    =\mathbf v_\theta(\mathbf x_t,t),
  \qquad \mathbf x_0\sim p_0.
  \label{eq:flow-ode}
\end{equation}
Conditional flow matching trains on source--target pairs
$(\mathbf x_0,\mathbf x_1)\sim\pi$. The coupling $\pi$ specifies how source
states are paired with users' target histories. A standard continuous path is
$\mathbf x_t=(1-t)\mathbf x_0+t\mathbf x_1$, with conditional velocity
$\mathbf x_1-\mathbf x_0$. With $t\sim\operatorname{Uniform}(0,1)$,
the model learns this velocity by minimizing
\begin{equation}
  \mathcal L_{\mathrm{CFM}}
    =\mathbb E_{t,(\mathbf x_0,\mathbf x_1)\sim\pi}
      \left[\|\mathbf v_\theta(\mathbf x_t,t)
        -(\mathbf x_1-\mathbf x_0)\|_2^2\right].
  \label{eq:conditional-flow-matching}
\end{equation}
Intermediate states are sampled directly from this path, so training does not
require solving the ODE. The coupling determines the states presented to the
model and their target velocities~\cite{liu2025flow}.

\subsection{Discrete Flow Reconstruction}
Continuous interpolation between binary endpoints can produce fractional
entries. FlowCF instead selects between the binary source $\mathbf x_{0,u}$
and target history $\mathbf x_{1,u}$ at each item coordinate~\cite{liu2025flow}:
\begin{equation}
  \mathbf x_{t,u}=\mathbf m_t\odot\mathbf x_{1,u}
    +(\mathbf 1-\mathbf m_t)\odot\mathbf x_{0,u},
  \label{eq:flowcf-interpolation}
\end{equation}
where each entry of $\mathbf m_t$ is independently sampled from
$\operatorname{Bernoulli}(t)$ and $\odot$ denotes element-wise multiplication.
The conditional mean interpolates linearly between the endpoints while each
sampled state remains binary. FlowCF adopts a data-prediction parameterization:
a time-conditioned predictor
$f_\theta(\mathbf x_{t,u},t)$ estimates the target history using the
reconstruction objective
\begin{equation}
  \mathcal L_{\mathrm{rec}}
  =\mathbb E\left[
    \frac{\|f_\theta(\mathbf x_{t,u},t)-\mathbf x_{1,u}\|_2^2}
         {|\mathcal I|}\right],
  \label{eq:sourceflow-reconstruction}
\end{equation}
with the expectation over users, sampled times, and training states.
CoFlowCF further regularizes hidden representations with a contrastive
loss~\cite{jin2026contrastive}.

\section{Methodology}
\label{sec:methodology}
SourceFlow constructs collaboratively coupled training states in FlowCF,
retaining its prediction network, objective, and inference procedure.
We first analyze how source--target coupling affects reconstruction
(Section~\ref{sec:source-coupling}). We then introduce three components.
Neighborhood Coupling builds displaced binary sources from collaborative
alternatives to the target history (Section~\ref{sec:neighborhood-coupling}).
Collaborative Consensus refines
replacement probabilities using broader collaborative support
(Section~\ref{sec:collaborative-consensus}). Stretched Restoration
keeps replacements at late flow times during training, so the model continues
to learn to recover displaced interactions (Section~\ref{sec:late-time-retention}).
Figure~\ref{fig:sourceflow-overview} summarizes training and inference from
observed histories.

\subsection{From History Corruption to Explicit Source Coupling}
\label{sec:source-coupling}

\paragraph{Source-dependent reconstruction.}
Intermediate states can retain target entries even when the source is sampled
independently. Nevertheless, a source rule $q$ changes their conditional
reconstruction problem. For a fixed interpolation scheme, define
\begin{equation}
  K_{q,t}(\mathbf z\mid\mathbf x)
    =\Pr_q(\mathbf X_t=\mathbf z\mid\mathbf X_1=\mathbf x,t).
  \label{eq:source-state-kernel}
\end{equation}
The target distribution $p_{\mathrm{data}}$ remains unchanged, but the
conditional distribution of histories to reconstruct satisfies
\begin{equation}
  p_q(\mathbf x\mid\mathbf z,t)
    \propto p_{\mathrm{data}}(\mathbf x)
      K_{q,t}(\mathbf z\mid\mathbf x).
  \label{eq:source-restoration-posterior}
\end{equation}
Consequently, the Bayes-optimal predictor for the reconstruction objective
in Eq.~\eqref{eq:sourceflow-reconstruction} is
\begin{equation}
  f_q^*(\mathbf z,t)
    =\mathbb E_q[\mathbf X_1\mid\mathbf X_t=\mathbf z,t].
  \label{eq:source-optimal-reconstruction}
\end{equation}
Figure~\ref{fig:source-rule-posterior} illustrates this dependence: different
source rules can favor different restoration targets for the same intermediate
state, even with unchanged target histories and loss. Source coupling therefore
shapes the restoration relation learned during training.

At $t=0$, let $R_q^*(0)$ denote the minimum reconstruction MSE in
Eq.~\eqref{eq:sourceflow-reconstruction} and
$\boldsymbol\mu=\mathbb E[\mathbf X_1]$. An independent global source has
optimal prediction $\boldsymbol\mu$. Conditional-expectation orthogonality gives
\begin{equation}
  \begin{aligned}
    &R_{\mathrm G}^*(0)-R_q^*(0)\\
    &\quad=\frac{1}{|\mathcal I|}\mathbb E_q
      \bigl\|f_q^*(\mathbf X_0,0)-\boldsymbol\mu\bigr\|_2^2\geq0.
  \end{aligned}
  \label{eq:source-reconstruction-gain}
\end{equation}
The reduction is strict when the conditional target mean varies with the
source. Using the target history itself as the source gives zero reconstruction
risk but leaves no restoration task.
A displaced source can still retain recoverable information through its
coupling with the target. Appendix~\ref{app:source-risk} gives the full
proof; experiments assess the benefit along the path and for recommendation.

\paragraph{From independent sources to collaborative coupling.}
Diffusion perturbations originate from the user's history and therefore remain
target-dependent~\cite{wang2023diffusion}. FlowCF instead draws a source from a
global frequency-based Bernoulli prior, yielding the independent coupling
$p_0(\mathbf x_0)p_{\mathrm{data}}(\mathbf x_1)$. This supplies displacement
but no source-side information from the paired history. SourceFlow combines
displacement with target dependence through
\begin{equation}
  \pi(\mathbf x_0,\mathbf x_1)
  =p_{\mathrm{data}}(\mathbf x_1)
    q_{\mathrm{collab}}(\mathbf x_0\mid\mathbf x_1).
  \label{eq:collaborative-coupling}
\end{equation}
Neighborhood sampling implements this coupling: the target history determines
the anchors and eligible replacements, and the resulting source is paired
with that same history for reconstruction. Thus, the sampled alternatives
define which displaced interactions the model must recover. Consensus refines
this conditional distribution, while Stretched Restoration controls how long
its replacements remain in the training states.

\subsection{Neighborhood Coupling}
\label{sec:neighborhood-coupling}
We use every historical item as an anchor, drawing on the user's full observed
profile. Replacements are sampled from each anchor's unobserved collaborative
neighbors, whose co-occurrence with the anchor provides cues for recovering
the displaced interaction.

We construct an item graph from training interactions, discard weakly
supported co-occurrences, and weight the remaining edges by cosine similarity.
Writing $\mathcal U_i=\{v:i\in\mathcal H_v\}$ for the users who interacted with
item $i$, the weight of a retained edge is
\begin{equation}
  w_{ij}=\frac{|\mathcal U_i\cap\mathcal U_j|}
                   {\sqrt{|\mathcal U_i|\,|\mathcal U_j|}}.
  \label{eq:collaborative-weight}
\end{equation}
The weight normalizes co-occurrence by item prevalence. Let
$\mathcal N_K(i)$ denote the retained top-$K$ neighbors of $i$.

The eligible pool $\mathcal C_u(i)=\mathcal N_K(i)\setminus\mathcal H_u$
contains the anchor's top-$K$ neighbors outside user $u$'s history.
We sample an eligible replacement according to
\begin{equation}
  q_L(j\mid i,\mathcal H_u)
  =\frac{w_{ij}}{\sum_{\ell\in\mathcal C_u(i)}w_{i\ell}},
  \quad j\in\mathcal C_u(i).
  \label{eq:local-coupling}
\end{equation}
If $\mathcal C_u(i)=\varnothing$, we draw the replacement from the item-popularity
distribution estimated from training data, excluding $\mathcal H_u$.
Sampling one replacement $j_{u,i}$ for each anchor gives a set of pairs
$\mathcal P_u=\{(i,j_{u,i}):i\in\mathcal H_u\}$ and the binary source
$\mathbf x_{0,u}=\mathbf b(\{j_{u,i}:i\in\mathcal H_u\})$.
The pair set preserves the slot-wise anchor--replacement correspondence used
by the training path in Section~\ref{sec:late-time-retention}.

\subsection{Collaborative Consensus}
\label{sec:collaborative-consensus}
Direct co-occurrence measures the association between an anchor and a
candidate, but does not summarize support from other items in histories
containing the anchor. Collaborative Consensus aggregates
this support to refine replacement probabilities within the same candidate
pool.

For a training user $v$ who interacted with $i$, we average the precomputed
cosine weights $w_{kj}$ from Eq.~\eqref{eq:collaborative-weight} over the remaining
history, then aggregate across all users who interacted with $i$:
\begin{align}
  c_v(i,j)
    &=\frac{1}{|\mathcal H_v|-1}
      \sum_{k\in\mathcal H_v\setminus\{i\}}w_{kj},\notag\\
  \bar c(i,j)
    &=\frac{1}{|\mathcal U_i|}\sum_{v\in\mathcal U_i}c_v(i,j).
  \label{eq:collaborative-consensus}
\end{align}
$\bar c(i,j)$ captures indirect support across users through
$i\rightarrow k\rightarrow j$, complementing the direct
$i\rightarrow j$ relation.

We combine the two forms of evidence by reweighting the local distribution:
\begin{equation}
  q_C(j\mid i,\mathcal H_u)
  \propto w_{ij}\bigl(\varepsilon+\bar c(i,j)\bigr)^\beta,
  \quad j\in\mathcal C_u(i),
  \label{eq:consensus-coupling}
\end{equation}
where $\varepsilon>0$ smooths the score and $\beta$ controls consensus strength.
The pool remains fixed: $w_{ij}$ preserves first-order relevance, while
$\bar c(i,j)$ reallocates probability toward candidates with broader support.
The statistic is precomputed from training histories and shared across users
for each anchor--candidate pair. To construct a user's source, we apply these
shared scores to the candidates eligible for that user's anchors.
Appendix~\ref{app:consensus-refinement} derives an optimization interpretation
of this reweighting and its effect on expected support.

\subsection{Stretched Restoration}
\label{sec:late-time-retention}
Neighborhood Coupling and consensus determine the alternatives in the source;
the schedule determines how long they remain visible during training. Under
the linear schedule, replacements survive with probability $1-t$ and become
rare at late flow times. We retain more of them so the predictor still
needs to recover displaced interactions.

For each user, we sample a time
$t\in\mathcal T=\{0,1/N,\ldots,(N-1)/N\}$ and draw an independent
selector for each anchor--replacement pair:
\begin{equation}
  \rho_\gamma(t)=1-(1-t)^\gamma,
  \qquad m_{u,i}\sim\operatorname{Bernoulli}\bigl(\rho_\gamma(t)\bigr).
  \label{eq:stretched-retention}
\end{equation}
Here $\gamma>0$ is the retention exponent. Each slot selects either the
historical item or its paired substitute,
preserving their correspondence during interpolation.
The resulting state is $\mathbf x_{t,u}=\mathbf b(\mathcal H_{t,u})$, where
\begin{equation}
  \begin{split}
    \mathcal H_{t,u}
      ={}&\{i\in\mathcal H_u:m_{u,i}=1\}\\
       &\cup\{j_{u,i}:i\in\mathcal H_u,\ m_{u,i}=0\}.
  \end{split}
  \label{eq:slot-interpolation}
\end{equation}
Because $\mathcal H_{t,u}$ is a set, repeated replacements activate one binary
coordinate whenever any corresponding slot selects that item.
When $\gamma=1$, each slot selects its original history item with probability
$t$, giving a linear schedule for slot selection.
For $0<\gamma<1$, replacements survive with probability $(1-t)^\gamma>1-t$
at every interior time, while the source and target endpoints remain unchanged.

\begin{table*}[t!]
\centering
\caption{Overall performance. Best results are bolded, second-best results
are underlined, and improvements are relative to CoFlowCF.
Results in this table are averaged over three random seeds.
Standard deviations for CoFlowCF and SourceFlow are reported in
Table~\ref{tab:seed-variability}.}
\label{tab:overall_performance}
\fontsize{7.5}{9}\selectfont
\setlength{\tabcolsep}{1.0pt}
\renewcommand{\arraystretch}{1.16}
\begin{tabular*}{\textwidth}{@{\extracolsep{\fill}}ll*{4}{c}@{\hspace{5pt}}*{6}{c}@{\hspace{5pt}}*{4}{c}@{}}
\toprule
\multirow{2}{*}{Dataset}
& \multirow{2}{*}{Metric}
& \multicolumn{4}{c}{Discriminative CF}
& \multicolumn{6}{c}{Generative CF}
& \multicolumn{3}{c}{Flow-based CF}
& \multirow{2}{*}{Improv.} \\
\cmidrule(lr){3-6}
\cmidrule(lr){7-12}
\cmidrule(lr){13-15}
&
& ItemKNN & MF-BPR & LightGCN & SimGCL
& CDAE & Mult-VAE & MacridVAE & RecVAE & DiffRec & GiffCF
& FlowCF & CoFlowCF & \textbf{SourceFlow}
& \\
\midrule
\multirow{4}{*}{Sports}
& R@10
& 0.0319 & 0.0426 & 0.0482 & 0.0561
& 0.0526 & 0.0484 & 0.0525 & 0.0539 & 0.0538 & 0.0395
& 0.0563 & \underline{0.0588} & \textbf{0.0724}
& +23.13\% \\
& R@20
& 0.0455 & 0.0651 & 0.0742 & 0.0818
& 0.0777 & 0.0763 & 0.0746 & 0.0790 & 0.0765 & 0.0587
& 0.0799 & \underline{0.0838} & \textbf{0.1048}
& +25.06\% \\
& N@10
& 0.0194 & 0.0237 & 0.0267 & 0.0312
& 0.0299 & 0.0264 & 0.0317 & 0.0302 & 0.0317 & 0.0237
& 0.0326 & \underline{0.0337} & \textbf{0.0406}
& +20.47\% \\
& N@20
& 0.0229 & 0.0294 & 0.0334 & 0.0378
& 0.0363 & 0.0335 & 0.0373 & 0.0367 & 0.0375 & 0.0286
& 0.0387 & \underline{0.0401} & \textbf{0.0488}
& +21.70\% \\
\midrule
\multirow{4}{*}{Toys}
& R@10
& 0.0819 & 0.0707 & 0.0780 & 0.0752
& 0.0756 & 0.0760 & 0.0798 & 0.0805 & 0.0835 & \underline{0.0986}
& 0.0836 & 0.0891 & \textbf{0.1100}
& +23.46\% \\
& R@20
& 0.1067 & 0.1003 & 0.1060 & 0.1000
& 0.1047 & 0.1066 & 0.1057 & 0.1076 & 0.1136 & \underline{0.1295}
& 0.1139 & 0.1218 & \textbf{0.1483}
& +21.76\% \\
& N@10
& 0.0502 & 0.0423 & 0.0464 & 0.0466
& 0.0465 & 0.0454 & 0.0502 & 0.0493 & 0.0505 & \underline{0.0630}
& 0.0507 & 0.0535 & \textbf{0.0658}
& +22.99\% \\
& N@20
& 0.0566 & 0.0499 & 0.0536 & 0.0529
& 0.0539 & 0.0532 & 0.0568 & 0.0562 & 0.0582 & \underline{0.0710}
& 0.0585 & 0.0619 & \textbf{0.0757}
& +22.29\% \\
\midrule
\multirow{4}{*}{Video}
& R@10
& 0.0953 & 0.1032 & 0.1139 & 0.1222
& 0.1066 & 0.1112 & 0.1132 & 0.1148 & 0.1204 & \underline{0.1336}
& 0.1230 & 0.1255 & \textbf{0.1415}
& +12.75\% \\
& R@20
& 0.1358 & 0.1505 & 0.1668 & 0.1667
& 0.1534 & 0.1653 & 0.1634 & 0.1685 & 0.1681 & \underline{0.1928}
& 0.1738 & 0.1795 & \textbf{0.2047}
& +14.04\% \\
& N@10
& 0.0534 & 0.0571 & 0.0632 & 0.0705
& 0.0611 & 0.0609 & 0.0638 & 0.0627 & 0.0695 & \underline{0.0762}
& 0.0701 & 0.0711 & \textbf{0.0792}
& +11.39\% \\
& N@20
& 0.0638 & 0.0693 & 0.0768 & 0.0819
& 0.0732 & 0.0747 & 0.0767 & 0.0764 & 0.0817 & \underline{0.0904}
& 0.0832 & 0.0850 & \textbf{0.0955}
& +12.35\% \\
\midrule
\multirow{4}{*}{Beauty}
& R@10
& 0.0981 & 0.0942 & 0.1049 & 0.0962
& 0.1074 & 0.1138 & 0.1156 & 0.1126 & 0.1167 & 0.1215
& 0.1229 & \underline{0.1264} & \textbf{0.1365}
& +7.99\% \\
& R@20
& 0.1346 & 0.1362 & 0.1417 & 0.1238
& 0.1526 & 0.1559 & 0.1573 & 0.1544 & 0.1562 & 0.1608
& 0.1638 & \underline{0.1670} & \textbf{0.1812}
& +8.50\% \\
& N@10
& 0.0596 & 0.0530 & 0.0581 & 0.0589
& 0.0625 & 0.0667 & 0.0712 & 0.0666 & 0.0714 & 0.0729
& 0.0745 & \underline{0.0764} & \textbf{0.0815}
& +6.68\% \\
& N@20
& 0.0692 & 0.0641 & 0.0677 & 0.0662
& 0.0743 & 0.0778 & 0.0821 & 0.0775 & 0.0817 & 0.0833
& 0.0853 & \underline{0.0872} & \textbf{0.0932}
& +6.88\% \\
\midrule
\multirow{4}{*}{Yelp}
& R@10
& 0.0329 & 0.0400 & 0.0426 & 0.0510
& 0.0491 & 0.0480 & 0.0478 & 0.0488 & 0.0465 & 0.0521
& 0.0511 & \underline{0.0547} & \textbf{0.0608}
& +11.15\% \\
& R@20
& 0.0504 & 0.0655 & 0.0700 & 0.0785
& 0.0745 & 0.0775 & 0.0753 & 0.0771 & 0.0716 & 0.0837
& 0.0799 & \underline{0.0853} & \textbf{0.0950}
& +11.37\% \\
& N@10
& 0.0189 & 0.0215 & 0.0232 & 0.0282
& 0.0277 & 0.0262 & 0.0268 & 0.0265 & 0.0268 & 0.0294
& 0.0292 & \underline{0.0306} & \textbf{0.0341}
& +11.44\% \\
& N@20
& 0.0236 & 0.0282 & 0.0305 & 0.0355
& 0.0345 & 0.0339 & 0.0340 & 0.0340 & 0.0335 & 0.0377
& 0.0368 & \underline{0.0387} & \textbf{0.0431}
& +11.37\% \\
\bottomrule
\end{tabular*}
\end{table*}

The resulting states train the predictor with
Eq.~\eqref{eq:sourceflow-reconstruction}. Source construction and stretching
apply only during training, teaching the predictor to recover missing
interactions from collaborative cues at late flow times. At inference, the
observed history likewise provides only partial evidence of the user's
preferences. Starting from this history at the same late flow times, we
follow the two-step rule of FlowCF and
CoFlowCF~\cite{liu2025flow,jin2026contrastive}: the first step updates the
binary state while retaining observed items, and the second draws on these learned
collaborative associations to rank unobserved items.
Algorithm~\ref{alg:sourceflow} summarizes training and inference.

\section{Experiments}
We organize the experiments around six research questions:
\begin{itemize}
  \item \textbf{RQ1:} How does SourceFlow perform compared with existing
  recommendation methods?
  \item \textbf{RQ2:} What gains do Neighborhood Coupling,
  Collaborative Consensus, and Stretched Restoration contribute?
  \item \textbf{RQ3:} How do collaborative source structure and source--target
  pairing affect reconstruction and recommendation?
  \item \textbf{RQ4:} How does Stretched Restoration affect recovery
  and test-set ranking on collaborative perturbation states?
  \item \textbf{RQ5:} How do the gains vary across item popularity levels?
  \item \textbf{RQ6:} How sensitive is SourceFlow to its key hyperparameters,
  and how do its training time per epoch and inference time compare with Base?
\end{itemize}

\subsection{Experimental Setup}
Unless otherwise specified, we follow the experimental protocol and
implementation settings of CoFlowCF~\cite{jin2026contrastive}.

\paragraph{Datasets.}
We evaluate on five real-world datasets: Sports, Toys, Video, Beauty,
and Yelp. For each user, we split interactions into training, validation,
and test sets with an 8:1:1 ratio.
Table~\ref{tab:dataset-statistics} in Appendix~\ref{app:implementation}
reports the dataset statistics.

\paragraph{Baselines.}
We compare with three groups of methods. Discriminative CF baselines include
ItemKNN \cite{sarwar2001item}, MF-BPR \cite{rendle2012bpr},
LightGCN \cite{he2020lightgcn}, and SimGCL \cite{yu2022graph}.
Generative baselines include CDAE \cite{wu2016collaborative},
Mult-VAE \cite{liang2018variational}, MacridVAE \cite{ma2019learning},
RecVAE \cite{shenbin2020recvae}, DiffRec \cite{wang2023diffusion}, and
GiffCF \cite{zhu2024graph}.
Flow-based baselines include FlowCF \cite{liu2025flow} and
CoFlowCF \cite{jin2026contrastive}.
In the ablation studies, \emph{Base} denotes CoFlowCF.

\paragraph{Evaluation Protocol.}
We report Recall and NDCG at cutoffs 10 and 20 under the full-ranking
protocol and select checkpoints by validation NDCG@10.

\paragraph{Implementation Details.}
The collaborative graph and consensus statistics use only training
interactions. We select $\gamma$ once per dataset by validation NDCG@10
and fix it across the three-seed runs. Implementation details are in
Appendix~\ref{app:implementation}.

\subsection{Overall Performance (RQ1)}
SourceFlow achieves substantial and consistent gains across all five datasets,
ranking first in all 20 dataset--metric combinations
(Table~\ref{tab:overall_performance}). Relative Recall@10 gains over CoFlowCF
range from 8.0\% to 23.5\%, averaging 15.7\%. On Sports and Toys,
improvements exceed 20\% across all four metrics.

These gains come from redesigning training sources and their retention
schedule, with CoFlowCF's prediction network and two-step inference procedure
unchanged. Sources carry collaborative cues from the paired history, and
stretch retains these cues at late flow times. The following ablations
separate the contributions of source construction, consensus, and stretch.

\subsection{Ablation Study (RQ2)}
\begin{table}[!ht]
  \centering
  \caption{Incremental component ablation on Sports and Yelp, starting from CoFlowCF (Base).}
  \label{tab:component-ablation}
  \normalsize
  \setlength{\tabcolsep}{2pt}
  \renewcommand{\arraystretch}{1.30}
    \begin{tabular*}{\columnwidth}{@{\extracolsep{\fill}}l*{4}{c}@{}}
      \toprule
      \multirow{2}{*}{Component}
        & \multicolumn{2}{c}{Sports}
        & \multicolumn{2}{c}{Yelp} \\
      \cmidrule(lr){2-3}\cmidrule(lr){4-5}
        & R@10 & N@10 & R@10 & N@10 \\
      \midrule
      Base
        & 0.0588 & 0.0337 & 0.0547 & 0.0306 \\
      + Neighborhood
        & 0.0678 & 0.0377 & 0.0573 & 0.0322 \\
      + Collaborative Consensus
        & 0.0704 & 0.0395 & 0.0588 & 0.0331 \\
      \textbf{SourceFlow}
        & \textbf{0.0724} & \textbf{0.0406}
        & \textbf{0.0608} & \textbf{0.0341} \\
      \bottomrule
    \end{tabular*}
\end{table}

Table~\ref{tab:component-ablation} shows gains at each stage of the cumulative
ablation. Replacing the global source with collaborative neighbors gives
the largest initial improvement, increasing
Recall@10 by 15.3\% on Sports and 4.8\% on Yelp.
Collaborative Consensus improves both datasets without changing the candidate
pool. Higher-order support therefore helps select replacements beyond what
direct co-occurrence provides. Retaining these replacements at later times
improves performance further on both datasets: the gains depend on both which
alternatives are sampled and how long the predictor encounters them during
training. See Appendix~\ref{app:full-component-ablation} for other datasets.

\begingroup
\subsection{Source--Target Coupling Analysis (RQ3)}
\label{sec:source-preference-preservation}

We compare source choices and pairings to understand why collaborative sources
improve recommendation. Further tests examine preference preservation and whether
predictors use these cues to reconstruct target histories.

\begin{table}[t]
  \centering
  \caption{Comparison of source construction strategies on Sports. NDCG@10 gains are relative to Global.}
  \label{tab:source-construction}
  \normalsize
  \setlength{\tabcolsep}{2pt}
  \renewcommand{\arraystretch}{1.062}
  \begin{tabular*}{\columnwidth}{@{\extracolsep{\fill}}lccc@{}}
    \toprule
    Source & R@10 & N@10 & $\Delta$N@10 \\
    \midrule
    Global & 0.0588 & 0.0337 & --- \\
    User History & 0.0530 & 0.0305 & $-9.5\%$ \\
    Masked History & 0.0558 & 0.0324 & $-3.9\%$ \\
    Gaussian-Corrupted History & 0.0211 & 0.0117 & $-65.3\%$ \\
    Uniform Swap & 0.0547 & 0.0322 & $-4.5\%$ \\
    Popularity-Matched Swap & 0.0601 & 0.0343 & +1.8\% \\
    \midrule
    Ours (Neighborhood) & 0.0678 & 0.0377 & +11.9\% \\
    Ours (+Collaborative Consensus) & \textbf{0.0704} & \textbf{0.0395} & \textbf{+17.2\%} \\
    Shuffled Pairing & 0.0547 & 0.0310 & $-8.0\%$ \\
    \bottomrule
  \end{tabular*}
\end{table}

\paragraph{Source Construction.}
Table~\ref{tab:source-construction} compares source choices on Sports under a
linear schedule ($\gamma=1$). User History copies the target history, leaving
no restoration task. Masked History removes some target interactions, while
Gaussian-Corrupted History adds Gaussian noise (Appendix~\ref{app:implementation}).
Both corrupt the original history signal without supplying collaborative
alternatives that carry user-preference cues for reconstruction. Uniform Swap replaces historical items one for one with
uniformly sampled unobserved alternatives;
Popularity-Matched Swap samples them in proportion to global training
frequencies. All underperform Collaborative Consensus.
These results show that displacement and popularity matching alone do not
account for the collaborative gain.
We then test whether the gains depend on pairing these sources with the histories
that guided their construction. Reassigning the same source vectors across users
lowers NDCG@10 to 0.0310 under matched interpolation
(Appendix~\ref{app:interpolation-control}). Using the same neighborhoods for hard-negative
regularization also fails to improve Global across the tested weights
(Appendix~\ref{app:hard-negative-control}), supporting the value of collaborative
source design.

\paragraph{Preference-Group Preservation.}
Figure~\ref{fig:preference-group-transfer} tests whether sources retain
users' preference information after observed items are replaced. We cluster
users' target histories into five preference groups in an SVD space. Each
source is mapped into the same space and assigned to the nearest group
center (Appendix~\ref{app:preference-group-probe}). Global's nearly identical
rows, concentrated in the largest preference group (Group~2), show that the
user's preference group has little effect on the
group assigned to the source. SourceFlow's higher diagonal values show
that more sources are assigned to the same preference group as their paired
histories. Collaborative replacement thus preserves group-level preference
cues across different item sets.

\begin{figure}[!t]
  \centering
  \includegraphics[width=\columnwidth,trim=9 2 2 2,clip]{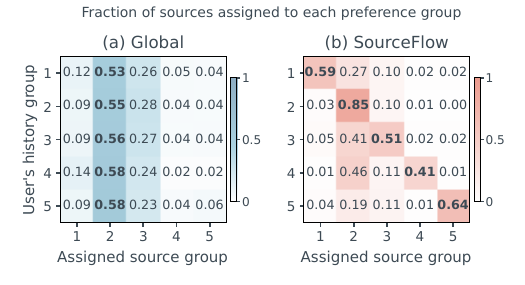}
  \caption{Preference-group preservation. Target histories are clustered into
  five groups in an SVD space. Sources are projected into this space and
  assigned to the nearest group center. SourceFlow's stronger diagonal (b)
  shows more matches to the paired users' groups than Global (a), indicating
  that replacements preserve group-level preference cues.}
  \Description{Two heatmaps compare Global and SourceFlow on Sports. Rows
  indicate users' history groups, and columns indicate the groups assigned to
  their sources. Cell values are fractions of users within each row.
  Diagonal cells represent matching source and history groups, with higher
  values for SourceFlow than Global.}
  \label{fig:preference-group-transfer}
\end{figure}

\paragraph{History Reconstruction.}
Figure~\ref{fig:paired-history-advantage} tests whether the preference cues
retained in sources (Figure~\ref{fig:preference-group-transfer}) help the predictor
reconstruct target histories. In panel (a), SourceFlow has lower MSE than Global
along the path, with the greatest reduction at the starting point ($t=0$).
At $t=0$ in panel (b), each prediction is compared with its paired target
history and another user's length-matched history. SourceFlow reconstructs
the paired history more accurately (lower MSE) for 92.4\% of users, compared with 50.0\%
for Global and 50.1\% for Shuffled Pairing
(Appendix~\ref{app:paired-history-probe}). Shuffling removes this advantage,
showing that user-specific
source--target coupling helps the predictor use the preserved preference cues.

\begin{figure}[!t]
  \centering
  \includegraphics[width=\columnwidth]{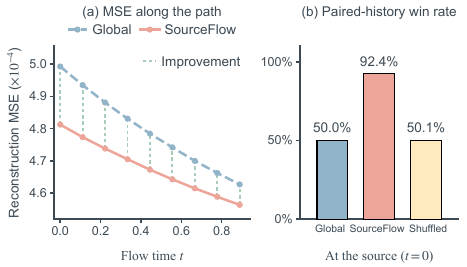}
  \caption{History reconstruction. (a) Target-history MSE along each method's
  flow path.
  (b) At $t=0$, the win rate is the percentage of users whose prediction has
  lower MSE for the paired target history than for an
  unpaired target history. SourceFlow uses
  Collaborative Consensus \mbox{($\gamma=1$)}.}
  \Description{Two panels compare history reconstruction. The left shows lower
  target-history MSE for SourceFlow than Global along the flow path, with the
  largest gap at the source endpoint. The right shows paired-history win rates
  of 50.0 percent for Global, 92.4 percent for SourceFlow, and 50.1 percent for
  Shuffled Pairing at time zero.}
  \label{fig:paired-history-advantage}
\end{figure}

\subsection{Restoration along the Training Path (RQ4)}
We compare models trained with Collaborative Consensus replacements under
linear ($\gamma=1$) and stretched ($\gamma=0.5$) schedules. At each flow time,
both receive the same intermediate states from a fixed stretched path on
Sports. We use each model's prediction for both panels. The left panel reports
pairwise recovery, the fraction of original historical items that the
predictor scores above the alternatives replacing them in the input. The right
reports test-set NDCG@10 from the same prediction.
Appendix~\ref{app:path-recovery-probe} gives details.

Figure~\ref{fig:path-recovery} shows higher pairwise recovery and test-set
NDCG@10 for the stretched model on shared states, with both advantages
persisting at late flow times. Retaining replacements trains the predictor
to score missing historical items using collaborative evidence. This ability
to score relevant items absent from the input also supports ranking
unobserved items from observed histories. Table~\ref{tab:component-ablation}
shows further recommendation gains from adding Stretched Restoration to
Collaborative Consensus.

\begin{figure}[!ht]
  \centering
  \includegraphics[width=\columnwidth]{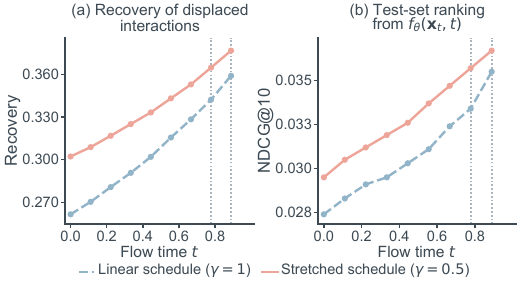}
  \caption{Restoration and recommendation along the flow path on Sports.
  On shared states, the stretched model scores original historical items
  above their replacements at a higher rate (a) and achieves higher test-set
  NDCG@10 (b).
  Dashed lines mark the flow times used by two-step inference,
  $t_a=(N-2)/N$ and $t_b=(N-1)/N$.}
  \Description{Two line charts on shared intermediate
  states compare linear and stretched schedules with Collaborative Consensus
  replacements. Dashed vertical lines mark two late flow times.}
  \label{fig:path-recovery}
\end{figure}
\par
\endgroup

\subsection{Performance across Item Popularity (RQ5)}
\label{sec:item-popularity}

\begin{figure}[!ht]
  \centering
  \includegraphics[width=0.49\columnwidth]{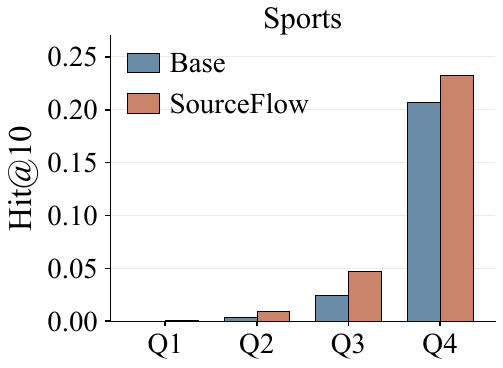}\hfill
  \includegraphics[width=0.49\columnwidth]{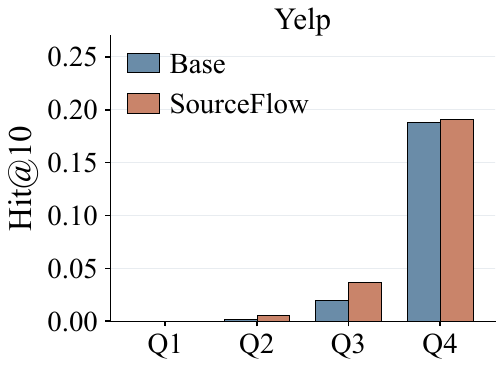}
  \par\medskip
  \includegraphics[width=0.49\columnwidth]{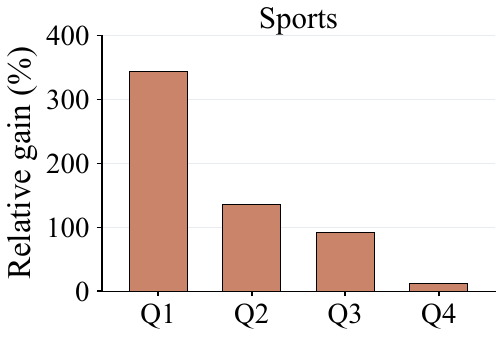}\hfill
  \includegraphics[width=0.49\columnwidth]{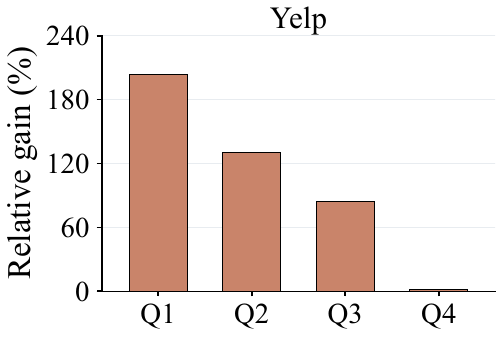}
  \caption{Popularity-stratified performance on Sports (left) and Yelp
  (right). Top: Hit@10; bottom: relative gains over Base. Q1 is the least
  popular quartile and Q4 the most popular.}
  \Description{Four panels compare Base and SourceFlow across item-popularity
  quartiles. The top row shows Hit at 10 and the bottom row shows relative
  gains over Base, with Sports on the left and Yelp on the right.}
  \label{fig:popularity-performance}
\end{figure}

We examine whether collaborative source coupling improves recommendation
performance on interactions beyond the most popular items. Test interactions
are divided into quartiles by their target items' training frequency, from the least popular
(Q1) to the most popular (Q4).
Figure~\ref{fig:popularity-performance} reports Hit@10 (the fraction of test
interactions ranked in the user's top 10) and relative gains over CoFlowCF.

SourceFlow improves Hit@10 across all quartiles on both datasets, with the largest
relative gains on less-popular items. The improvements in Q2--Q4 show that
the benefit extends beyond the tail
while retaining strong performance on the head.

The global prior captures item frequencies but offers
no source-side cue for a user's less-common interests. SourceFlow selects
replacements through historical anchors and broader collaborative support,
so the predictor learns to recover interactions with less-popular items from alternatives
in the same collaborative region rather than from popularity-driven states.
This mechanism is consistent with the larger relative gains on less-popular items.

\subsection{Parameter Sensitivity and Efficiency (RQ6)}

\begin{figure}[H]
  \centering
  \includegraphics[width=\columnwidth]{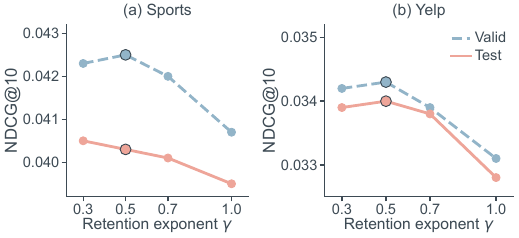}
  \caption{Stretch sensitivity on Sports and Yelp. Outlined markers indicate
  the validation-selected exponent.}
  \Description{Validation and test NDCG at 10 for retention exponents
  0.3, 0.5, 0.7, and 1.0 on Sports and Yelp, shown side by side.
  Validation performance peaks at 0.5 on both datasets.}
  \label{fig:stretch-sensitivity}
\end{figure}

\paragraph{Retention Schedule.}
Figure~\ref{fig:stretch-sensitivity} varies the retention exponent $\gamma$,
which controls how long replacements remain along the path. On both datasets,
$\gamma$ between
0.3 and 0.7 improves validation and test NDCG@10 over the linear schedule
($\gamma=1$), with validation performance peaking at $\gamma=0.5$.

\paragraph{Neighborhood Size.}
Neighborhood Coupling and Collaborative Consensus use $K=50$ neighbors per item
to limit storage. On Sports, $K=100$ and $200$ yield NDCG@10 of 0.0384 and
0.0380, respectively, both above Base (0.0337).
Overall, collaborative source design yields consistent gains across the tested
settings without requiring fine-grained parameter tuning.

\begin{table}[t]
  \centering
  \caption{Efficiency on Sports and Yelp.}
  \label{tab:efficiency}
  \small
  \setlength{\tabcolsep}{3pt}
  \renewcommand{\arraystretch}{1.12}
  \begin{tabular}{llccc}
    \toprule
    Dataset & Method & Params & Train (s/epoch) & Inference (s) \\
    \midrule
    Sports & Base      & 7.67M  & 0.12 & 0.33 \\
    Sports & SourceFlow & 7.67M  & 0.14 & 0.32 \\
    Yelp   & Base      & 39.14M & 3.09 & 4.21 \\
    Yelp   & SourceFlow & 39.14M & 3.09 & 4.22 \\
    \bottomrule
  \end{tabular}
\end{table}

\paragraph{Efficiency.}
SourceFlow retains Base's parameter count and full-sort inference procedure.
The graph and consensus scores are computed once before training.
Table~\ref{tab:efficiency} shows similar per-epoch training and full-sort
inference times for the two methods.

\section{Conclusion}
We proposed SourceFlow to couple training sources with user histories in
flow-based recommendation. It builds displaced binary sources by replacing
observed items with unobserved collaborative alternatives.
Neighborhood Coupling and Collaborative Consensus guide replacement
probabilities. Stretched Restoration keeps these replacements at late flow
times to help the predictor reconstruct the original history. Experiments on
five datasets demonstrate consistent recommendation gains with the prediction
network and inference procedure unchanged. Controlled analyses show that these
gains depend on pairing collaborative alternatives with the corresponding
histories. Source design can therefore improve history reconstruction by
providing collaborative cues beyond the history retained in intermediate states.

\bibliographystyle{ACM-Reference-Format}
\bibliography{sample-base}

\clearpage
\appendix

\section{SourceFlow Algorithm}
\label{app:sourceflow-algorithm}
Algorithm~\ref{alg:sourceflow} outlines the construction and use of training
states. Here $\Pi_{\mathcal H_u}$ thresholds each coordinate at $1/2$ and
retains the observed items in $\mathcal H_u$.

\begin{algorithm}[H]
  \caption{SourceFlow training and recommendation}
  \label{alg:sourceflow}
  \begin{algorithmic}
    \Statex \textbf{Input:} Target histories $\{\mathcal H_u\}$, predictor
      $f_\theta$, time grid $\mathcal T$, retention exponent $\gamma$.
    \Statex \textbf{Output:} Trained predictor and recommendation scores.
    \Function{Train}{ }
      \State Build item weights $W$ and neighborhoods $\mathcal N_K$
      \State Precompute $\bar c(i,j)$ by Eq.~\eqref{eq:collaborative-consensus}
      \For{each training step}
        \State Sample a mini-batch of users $\mathcal B$
        \ForAll{$u\in\mathcal B$}
          \State $\mathbf x_{1,u}\gets\mathbf b(\mathcal H_u)$;
            $t_u\sim\operatorname{Uniform}(\mathcal T)$
          \State Sample $j_{u,i}\sim q_C(\cdot\mid i,\mathcal H_u)$
            for each $i\in\mathcal H_u$
          \State Draw slot selectors by Eq.~\eqref{eq:stretched-retention}
          \State Construct $\mathbf x_{t_u,u}$ by Eq.~\eqref{eq:slot-interpolation}
        \EndFor
        \State Predict histories $f_\theta(\mathbf x_{t_u,u},t_u)$
        \State Update $\theta$ with $\mathcal L_{\mathrm{rec}}$ and inherited
          regularization
      \EndFor
      \State \Return $\theta$
    \EndFunction
    \Statex
    \Function{Recommend}{$\mathcal H_u,\theta$}
      \State $\mathbf x\gets\mathbf b(\mathcal H_u)$
      \State $t_a\gets(N-2)/N$; $t_b\gets(N-1)/N$
      \State $\hat{\mathbf x}_1\gets f_\theta(\mathbf x,t_a)$
      \State $\mathbf x\gets\Pi_{\mathcal H_u}\!\left(
        \mathbf x+\dfrac{t_b-t_a}{1-t_a}(\hat{\mathbf x}_1-\mathbf x)\right)$
      \State \Return $f_\theta(\mathbf x,t_b)$ for ranking unobserved items
    \EndFunction
  \end{algorithmic}
\end{algorithm}

\section{Implementation Details}
\label{app:implementation}

\begin{table}[H]
  \centering
  \caption{Dataset statistics after preprocessing.}
  \label{tab:dataset-statistics}
  \normalsize
  \setlength{\tabcolsep}{3.2pt}
  \renewcommand{\arraystretch}{1.15}
  \begin{tabular*}{\columnwidth}{@{\extracolsep{\fill}}lrrrr@{}}
    \toprule
    Dataset & \#Users & \#Items & \#Interactions & Density \\
    \midrule
    Sports & 22,686 & 12,302 & 185,718 & 0.07\% \\
    Toys & 11,268 & 7,310 & 95,420 & 0.12\% \\
    Video & 14,494 & 6,951 & 132,209 & 0.13\% \\
    Beauty & 10,553 & 6,087 & 94,148 & 0.15\% \\
    Yelp & 136,346 & 64,670 & 1,856,992 & 0.02\% \\
    \bottomrule
  \end{tabular*}
\end{table}

\paragraph{Evaluation Protocol.}
Consistent with all compared methods, we study non-sequential top-$N$
collaborative filtering: each user history is an unordered set of observed
interactions, equivalently represented as a binary vector, and interaction
order is not modeled. We report Recall@$K$
(R@$K$) and NDCG@$K$ (N@$K$), with
$K\in\{10,20\}$. Under the full-ranking protocol, we rank all unobserved
items for each user. Each user's interactions are randomly split into
training, validation, and test sets at an 8:1:1 ratio.

\paragraph{Training Configuration.}
We implement SourceFlow in PyTorch with RecBole and optimize it using Adam
with a learning rate of $10^{-3}$, a batch size of 4,096, and no weight decay.
Training runs for at most 500 epochs, with validation after each epoch and
early stopping after 20 epochs without improvement in validation NDCG@10.
We restore the checkpoint with the highest validation NDCG@10 for testing.
Table~\ref{tab:training-configuration} lists dataset-specific settings;
$\lambda$ and $\tau$ denote the weight and temperature of the inherited
contrastive loss. Across datasets, the initial dropout rate is zero and the
time-embedding dimension is 10. Following CoFlowCF's discretization-step analysis,
we use $N=9$ and retain its two-step inference procedure~\cite{jin2026contrastive}.
Within each dataset, Base and SourceFlow use the same hidden sizes, hidden
dropout, $\lambda$, and $\tau$; only the source construction and retention
schedule differ.

\begin{table}[H]
  \centering
  \caption{Dataset-specific settings shared by Base and SourceFlow.}
  \label{tab:training-configuration}
  \normalsize
  \setlength{\tabcolsep}{4pt}
  \renewcommand{\arraystretch}{1.1}
  \begin{tabular*}{\columnwidth}{@{\extracolsep{\fill}}lcccc@{}}
    \toprule
    Dataset & Hidden sizes & Dropout & $\lambda$ & $\tau$ \\
    \midrule
    Sports & [300, 300] & 0.10 & $10^{-3}$ & 0.5 \\
    Toys   & [300, 300] & 0.10 & $10^{-4}$ & 0.2 \\
    Video  & [300, 300] & 0.10 & $5\!\times\!10^{-4}$ & 0.2 \\
    Beauty & [600]      & 0.10 & $5\!\times\!10^{-3}$ & 0.5 \\
    Yelp   & [300, 300] & 0.05 & $10^{-4}$ & 0.5 \\
    \bottomrule
  \end{tabular*}
\end{table}

\paragraph{Stretch Search.}
We search $\gamma$ over $\{0.3,0.5,0.7,1.0\}$ using single-seed validation
NDCG@10 and fix the selected value across the three-seed runs.
We select $\gamma=0.5$ on Sports, Toys, Video, and Yelp, and $\gamma=1.0$
on Beauty.
The candidate pool size $K=50$, consensus exponent $\beta=1$,
and smoothing constant $\varepsilon=10^{-3}$ remain fixed in this search.

\paragraph{Source Construction.}
We discard item pairs that co-occur in fewer than two users' training
histories before cosine normalization and retain the top 50 neighbors of each
anchor item. The graph, consensus scores, and fallback popularity distribution
are computed once from training interactions. Replacements are resampled for
each target history, excluding its observed items. If an anchor has no eligible
neighbor, we draw from the training-item
popularity distribution, again excluding the history.
Replacement and retention operate slot by slot; repeated selections of an
item map to a single active coordinate in the binary state.

For Masked History, each target interaction is retained independently with
probability 0.5. For Gaussian-Corrupted History, we add i.i.d. $\mathcal N(0,1)$
noise to the target history.

\section{Source-Coupling Storage}
\label{app:coupling-storage}
The source sampler stores three arrays indexed by item and neighbor, each of size
$|\mathcal I|\times 50$: neighbor indices (\texttt{sim\_idx}), log weights
(\texttt{sim\_logw}), and consensus scores (\texttt{mean\_c}).
Table~\ref{tab:coupling-storage} reports their storage footprint, which is
smaller than the network weights on both datasets. These arrays are used
for training-state construction and are not required for inference.

\begin{table}[!htbp]
  \centering
  \caption{Source-coupling storage and network weight size (MB, rounded).}
  \label{tab:coupling-storage}
  \normalsize
  \setlength{\tabcolsep}{4pt}
  \renewcommand{\arraystretch}{1.1}
  \begin{tabular*}{\columnwidth}{@{\extracolsep{\fill}}lrr@{}}
    \toprule
    Stored component & Sports & Yelp \\
    \midrule
    Neighbor indices & 4.7 & 24.7 \\
    Log weights & 2.4 & 12.3 \\
    Consensus scores & 2.4 & 12.3 \\
    \midrule
    Total coupling storage & 9.4 & 49.3 \\
    Network weights & 29 & 149 \\
    \bottomrule
  \end{tabular*}
\end{table}

\section{Full Component Ablation}
\label{app:full-component-ablation}
Table~\ref{tab:full-component-ablation} extends the component analysis to the
three datasets not shown in Table~\ref{tab:component-ablation}. Neighborhood
Coupling provides the largest initial gain, and Consensus improves all three
datasets. Stretch adds further gains on Toys and Video. On Beauty, validation
selects $\gamma=1$, so the full method retains the linear schedule.

\begin{table}[!htbp]
  \centering
  \caption{Component ablation on Toys, Video, and Beauty.}
  \label{tab:full-component-ablation}
  \small
  \setlength{\tabcolsep}{2.2pt}
  \renewcommand{\arraystretch}{1.12}
  \begin{tabular*}{\columnwidth}{@{\extracolsep{\fill}}l*{6}{c}@{}}
    \toprule
    \multirow{2}{*}{Component}
      & \multicolumn{2}{c}{Toys}
      & \multicolumn{2}{c}{Video}
      & \multicolumn{2}{c}{Beauty} \\
    \cmidrule(lr){2-3}\cmidrule(lr){4-5}\cmidrule(lr){6-7}
      & R@10 & N@10 & R@10 & N@10 & R@10 & N@10 \\
    \midrule
    Base
      & 0.0891 & 0.0535 & 0.1255 & 0.0711 & 0.1264 & 0.0764 \\
    + Neighborhood
      & 0.1057 & 0.0631 & 0.1363 & 0.0770 & 0.1347 & 0.0811 \\
    + Consensus
      & 0.1086 & 0.0647 & 0.1400 & 0.0786 & \textbf{0.1365} & \textbf{0.0815} \\
    \textbf{SourceFlow}
      & \textbf{0.1100} & \textbf{0.0658}
      & \textbf{0.1415} & \textbf{0.0792}
      & \textbf{0.1365} & \textbf{0.0815} \\
    \bottomrule
\end{tabular*}
\end{table}

\section{Performance across Random Seeds}
\label{app:seed-variability}
Table~\ref{tab:seed-variability} reports the mean and standard deviation
across three random seeds for CoFlowCF and SourceFlow on all five datasets.

\begin{table}[!htbp]
  \centering
  \caption{Performance across three random seeds. Each cell shows the mean
  with $\pm$ standard deviation below.}
  \label{tab:seed-variability}
  \normalsize
  \setlength{\tabcolsep}{3pt}
  \renewcommand{\arraystretch}{1.15}
  \begin{tabular*}{\columnwidth}{@{\extracolsep{\fill}}lcccc@{}}
    \toprule
    Method & R@10 & R@20 & N@10 & N@20 \\
    \midrule
    \multicolumn{5}{c}{\textit{Sports}} \\
    \midrule
    CoFlowCF & \shortstack{0.0588\\$\pm 0.0031$} & \shortstack{0.0838\\$\pm 0.0042$} & \shortstack{0.0337\\$\pm 0.0017$} & \shortstack{0.0401\\$\pm 0.0020$} \\
    SourceFlow & \shortstack{0.0724\\$\pm 0.0012$} & \shortstack{0.1048\\$\pm 0.0006$} & \shortstack{0.0406\\$\pm 0.0007$} & \shortstack{0.0488\\$\pm 0.0004$} \\
    \midrule
    \multicolumn{5}{c}{\textit{Toys}} \\
    \midrule
    CoFlowCF & \shortstack{0.0891\\$\pm 0.0016$} & \shortstack{0.1218\\$\pm 0.0006$} & \shortstack{0.0535\\$\pm 0.0006$} & \shortstack{0.0619\\$\pm 0.0003$} \\
    SourceFlow & \shortstack{0.1100\\$\pm 0.0023$} & \shortstack{0.1483\\$\pm 0.0019$} & \shortstack{0.0658\\$\pm 0.0017$} & \shortstack{0.0757\\$\pm 0.0013$} \\
    \midrule
    \multicolumn{5}{c}{\textit{Video}} \\
    \midrule
    CoFlowCF & \shortstack{0.1255\\$\pm 0.0009$} & \shortstack{0.1795\\$\pm 0.0012$} & \shortstack{0.0711\\$\pm 0.0008$} & \shortstack{0.0850\\$\pm 0.0009$} \\
    SourceFlow & \shortstack{0.1415\\$\pm 0.0037$} & \shortstack{0.2047\\$\pm 0.0030$} & \shortstack{0.0792\\$\pm 0.0018$} & \shortstack{0.0955\\$\pm 0.0016$} \\
    \midrule
    \multicolumn{5}{c}{\textit{Beauty}} \\
    \midrule
    CoFlowCF & \shortstack{0.1264\\$\pm 0.0011$} & \shortstack{0.1670\\$\pm 0.0004$} & \shortstack{0.0764\\$\pm 0.0001$} & \shortstack{0.0872\\$\pm 0.0002$} \\
    SourceFlow & \shortstack{0.1365\\$\pm 0.0033$} & \shortstack{0.1812\\$\pm 0.0018$} & \shortstack{0.0815\\$\pm 0.0021$} & \shortstack{0.0932\\$\pm 0.0018$} \\
    \midrule
    \multicolumn{5}{c}{\textit{Yelp}} \\
    \midrule
    CoFlowCF & \shortstack{0.0547\\$\pm 0.0005$} & \shortstack{0.0853\\$\pm 0.0002$} & \shortstack{0.0306\\$\pm 0.0003$} & \shortstack{0.0387\\$\pm 0.0002$} \\
    SourceFlow & \shortstack{0.0608\\$\pm 0.0003$} & \shortstack{0.0950\\$\pm 0.0002$} & \shortstack{0.0341\\$\pm 0.0003$} & \shortstack{0.0431\\$\pm 0.0002$} \\
    \bottomrule
  \end{tabular*}
\end{table}

\section{Analysis of Source Coupling}
\label{app:source-coupling-analysis}

\subsection{Reconstruction at the Source Endpoint}
\label{app:source-risk}
Fix the target distribution $p_{\mathrm{data}}$ and a source rule
$q(\mathbf x_0\mid\mathbf x_1)$. The minimum reconstruction MSE at $t=0$ is
\begin{equation}
  R_q^*(0)=\min_f\frac{1}{|\mathcal I|}
    \mathbb E_q\|\mathbf X_1-f(\mathbf X_0)\|_2^2,
  \label{eq:appendix-source-risk}
\end{equation}
where the minimum ranges over all predictors of the binary target vector.
Writing $m_q(\mathbf X_0)=\mathbb E_q[\mathbf X_1\mid\mathbf X_0]$, the
conditional residual has zero mean, so for any predictor $f$,
\begin{equation}
  \begin{aligned}
    &\mathbb E_q\|\mathbf X_1-f(\mathbf X_0)\|_2^2\\
    &\quad=\mathbb E_q\|\mathbf X_1-m_q(\mathbf X_0)\|_2^2\\
    &\qquad+\mathbb E_q\|m_q(\mathbf X_0)-f(\mathbf X_0)\|_2^2.
  \end{aligned}
  \label{eq:appendix-conditional-orthogonality}
\end{equation}
Hence $m_q$ attains the minimum. Under independent global sampling, the
conditional mean is the constant $\boldsymbol\mu=\mathbb E[\mathbf X_1]$,
and $R_{\mathrm G}^*(0)=\mathbb E\|\mathbf X_1-\boldsymbol\mu\|_2^2/|\mathcal I|$.
Taking $f=\boldsymbol\mu$ in
Eq.~\eqref{eq:appendix-conditional-orthogonality} and using the common target
distribution gives
\begin{equation}
  \begin{aligned}
    R_{\mathrm G}^*(0)
    &=R_q^*(0)\\
    &\quad+\frac{1}{|\mathcal I|}\mathbb E_q
      \|m_q(\mathbf X_0)-\boldsymbol\mu\|_2^2,
  \end{aligned}
\end{equation}
which proves Eq.~\eqref{eq:source-reconstruction-gain}. The gain is strictly
positive exactly when $m_q(\mathbf X_0)\neq\boldsymbol\mu$ with positive
probability.

\subsection{Consensus as Regularized Source Refinement}
\label{app:consensus-refinement}
Fix an anchor $i$, a history $\mathcal H_u$, and a nonempty eligible pool
$\mathcal C_u(i)$. Suppress these conditions in $q_L$ and $q_C$, and write
$c_j=\bar c(i,j)$ and $s_j=\log(\varepsilon+c_j)$.
The score $c_j$ is the expected collaborative support from a remaining
historical item. The expectation averages uniformly over users who interacted
with $i$ and over the other items in each user's history. It measures
compatibility with the interaction patterns surrounding the anchor.

For $\beta>0$, consider distributions $q$ over the same pool and the objective
\begin{equation}
  \mathcal J(q)=\mathbb E_{j\sim q}[s_j]
    -\frac{1}{\beta}D_{\mathrm{KL}}(q\|q_L).
  \label{eq:consensus-variational-objective}
\end{equation}
The first term favors broader collaborative support, while the second
penalizes departure from the local distribution. The normalized rule in
Eq.~\eqref{eq:consensus-coupling} is
\begin{equation*}
  q_C(j)=\frac{q_L(j)e^{\beta s_j}}{Z_\beta},
  \qquad Z_\beta=\sum_j q_L(j)e^{\beta s_j}.
\end{equation*}
Substitution yields
\begin{equation}
  \mathcal J(q)=\frac{1}{\beta}\log Z_\beta
    -\frac{1}{\beta}D_{\mathrm{KL}}(q\|q_C).
  \label{eq:consensus-variational-proof}
\end{equation}
Nonnegativity of KL divergence makes $q_C$ the unique maximizer.

For the setting $\beta=1$, the improvement in expected support also has a
closed form. Let $a=\mathbb E_{q_L}[c_j]$. Then
\begin{equation}
  \mathbb E_{q_C}[c_j]
    =\frac{\varepsilon a+\mathbb E_{q_L}[c_j^2]}{\varepsilon+a},
\end{equation}
and subtracting $a$ gives
\begin{equation}
  \mathbb E_{q_C}[c_j]-\mathbb E_{q_L}[c_j]
    =\frac{\operatorname{Var}_{q_L}(c_j)}{\varepsilon+a}\geq0.
  \label{eq:consensus-support-gain}
\end{equation}
Consensus therefore increases the expected higher-order support of each
sampled replacement whenever the candidates differ in that support. If all
scores coincide, $q_C=q_L$. The result characterizes the sampler's
collaborative compatibility; its effect on reconstruction and recommendation is
evaluated empirically.

\subsection{Preference-Group Probe}
\label{app:preference-group-probe}
For Figure~\ref{fig:preference-group-transfer}, we randomly sample 3,000 Sports users
with at least three training interactions. A rank-64 truncated
SVD of the training interaction matrix yields item factors $E$. Both
target histories and sources are mapped to $\phi(\mathbf x)=\operatorname{normalize}(\mathbf x E)$
using $\ell_2$ normalization. We cluster the history representations into five
groups using $k$-means.
The resulting group centers are fixed for both methods; each source is
assigned to its nearest center by Euclidean distance. We use one source draw
per user, with SourceFlow using Collaborative Consensus sources at $t=0$.
Heatmap rows are normalized within history groups and displayed
as fractions rounded to two decimal places. The user-weighted overall match
rates are 34.0\% for Global and 67.6\% for SourceFlow. Independent pairing under the
observed group marginals yields approximately 34\% for either source.

\subsection{History Reconstruction}
\label{app:paired-history-probe}
For Figure~\ref{fig:paired-history-advantage}, we use the trained Global,
Collaborative Consensus ($\gamma=1$, labelled SourceFlow), and Shuffled Pairing
checkpoints on Sports. Each variant uses its own predictor and source sampler.
Shuffled Pairing uses degree-matched source reassignment and a separately
trained predictor.

\paragraph{Path MSE}
For panel (a), we evaluate Global and SourceFlow on states from their
respective source and interpolation rules at $t=0,1/9,\ldots,8/9$.
We draw one state per user at each time and report MSE averaged over all users
and item coordinates, with predictors in evaluation mode. Pale green dashed
lines (Improvement) connect corresponding points, showing Global's MSE minus
SourceFlow's MSE.

\paragraph{Paired-history win rate.}
For panel (b), we fix another user $v(u)$ for each user $u$ by pairing neighboring users
after sorting by target-history length. This comparison pairing is shared
across variants. With $D=4$ source draws, let
$\hat{\mathbf x}_u^{(r)}=f_\theta(\mathbf x_{0,u}^{(r)},0)$ and define
\begin{equation}
  \begin{aligned}
    \Delta_u=\frac{1}{D|\mathcal I|}\sum_{r=1}^{D}
    \Bigl(&\|\hat{\mathbf x}_u^{(r)}-\mathbf x_{1,v(u)}\|_2^2\\
          &-\|\hat{\mathbf x}_u^{(r)}-\mathbf x_{1,u}\|_2^2\Bigr).
  \end{aligned}
  \label{eq:paired-history-advantage}
\end{equation}
Positive $\Delta_u$ means that the prediction is closer to the paired
target history. Errors are averaged over all item coordinates.
The paired-history win rate is the percentage of users with $\Delta_u>0$.

\subsection{Path-Recovery Probe}
\label{app:path-recovery-probe}
We draw one Collaborative Consensus replacement $j_{u,i}$ and one
$r_{u,i}\sim\operatorname{Uniform}(0,1)$ for every history slot. At each time,
both predictors receive the same state from
$m_{u,i}(t)=\mathbf 1[r_{u,i}\leq\rho_{0.5}(t)]$ and
Eq.~\eqref{eq:slot-interpolation}. Let
$\mathcal D_t=\{(u,i,j_{u,i}):m_{u,i}(t)=0\}$ be the displaced slot pairs.
For predictor $f_\theta$, we compute
\begin{equation}
  \operatorname{Recovery}_\theta(t)
  =\frac{1}{|\mathcal D_t|}
    \sum_{(u,i,j)\in\mathcal D_t}
    \mathbf 1\!\left[
      f_\theta(\mathbf x_{t,u},t)_i
      >f_\theta(\mathbf x_{t,u},t)_j
    \right].
  \label{eq:path-recovery-probe}
\end{equation}
Thus, a recovered pair means that the predictor ranks the displaced history
item above the substitute occupying its slot.
NDCG@10 is evaluated on test interactions, masking training and validation items.

\subsection{Interpolation and Pairing Control}
\label{app:interpolation-control}
We compare interpolation rules on Sports at $\gamma=1$, keeping the Consensus
source sampler and training settings fixed. Slot-wise
interpolation uses one selector per anchor--replacement pair, whereas
coordinate-wise interpolation independently selects an endpoint at each item
coordinate as in Eq.~\eqref{eq:flowcf-interpolation}. NDCG@10 is 0.0395 and
0.0394, respectively. Both the unshuffled and shuffled controls use the
coordinate-wise rule; shuffling reassigns source vectors between users of
similar history length and lowers NDCG@10 to 0.0310. Thus, the pairing change
has a much larger effect than the interpolation change in this comparison.

\section{Hard-Negative Control}
\label{app:hard-negative-control}
We keep the Global source and sample an eligible Top-50 neighbor $j$ as a
hard negative for each observed item $i$. We train with
$\mathcal L=\mathcal L_{\mathrm{rec}}+\lambda_{\mathrm{HN}}\operatorname{softplus}(s_j-s_i)$.
Here $s_i$ and $s_j$ are the predictor's item scores.
Table~\ref{tab:hard-negative-control} shows that no tested weight improves over
Base (N@10 0.0337), while SourceFlow reaches 0.0406. Thus, the neighborhood
graph does not reproduce the coupling gain when used only as a ranking
regularizer.

\begin{table}[!htbp]
  \centering
  \caption{Hard-negative control on Sports. $\Delta$ is relative to Base N@10.}
  \label{tab:hard-negative-control}
  \small
  \setlength{\tabcolsep}{2.1pt}
  \renewcommand{\arraystretch}{1.08}
  \begin{tabular*}{\columnwidth}{@{\extracolsep{\fill}}ccccccc@{}}
    \toprule
    $\lambda_{\mathrm{HN}}$ & Valid N@10 & R@10 & R@20 & N@10 & N@20 & $\Delta$ \\
    \midrule
    0.005 & 0.0234 & 0.0381 & 0.0544 & 0.0234 & 0.0275 & $-30.6\%$ \\
    0.01  & 0.0217 & 0.0368 & 0.0504 & 0.0221 & 0.0256 & $-34.4\%$ \\
    0.02  & 0.0206 & 0.0342 & 0.0467 & 0.0208 & 0.0240 & $-38.3\%$ \\
    0.1   & 0.0176 & 0.0299 & 0.0416 & 0.0177 & 0.0207 & $-47.5\%$ \\
    1.0   & 0.0104 & 0.0203 & 0.0324 & 0.0106 & 0.0136 & $-68.5\%$ \\
    \bottomrule
  \end{tabular*}
\end{table}

\end{document}